\documentclass[preprint,showpacs,preprintnumbers,amsmath,amssymb]{revtex4}

\usepackage{graphicx}
\usepackage{dcolumn}
\usepackage{bm}
\usepackage{array}

\begin{document}


 
\title{Cohesion Energetics of Carbon Allotropes : Quantum Monte Carlo Study}
\author{Hyeondeok Shin}
\author{Sinabro Kang}
\author{Jahyun Koo}
\author{Hoonkyung Lee}
\affiliation{
Division of Quantum Phases and Devices, School of Physics,
Konkuk University, Seoul 143-701, Korea}
\author{Jeongnim Kim}
 \email{jnkim@ornl.gov}
\affiliation{
Materials Science and Technology Division, Oak Ridge National Laboratory, TN 37831, U.S.A.
}
\author{Yongkyung Kwon}
 \email{ykwon@konkuk.ac.kr}
\affiliation{
Division of Quantum Phases and Devices, School of Physics,
Konkuk University, Seoul 143-701, Korea}

\date{\today}


\begin{abstract}
{We have performed quantum Monte Carlo calculations to study the cohesion energetics of carbon allotropes,
including $sp^3$-bonded diamond, $sp^2$-bonded graphene, $sp$-$sp^2$ hybridized graphynes, and $sp$-bonded
carbyne. The computed cohesive energies of diamond and graphene are found to be in excellent agreement 
with the corresponding values determined experimentally for diamond and graphite, respectively, 
when the zero-point energies, along with the interlayer binding in the case of graphite, are included. 
We have also found that the cohesive energy of graphyne decreases systematically as the ratio 
of $sp$-bonded carbon atoms increases. The cohesive energy of $\gamma$-graphyne, 
the most energetically-stable graphyne, turns out to be 6.766(6) eV/atom, which is smaller 
than that of graphene by 0.698(12) eV/atom. Experimental difficulty in synthesizing graphynes 
could be explained by their significantly smaller cohesive energies. 
Finally we conclude that the cohesive energy of a newly-proposed graphyne can be accurately estimated 
with the carbon-carbon bond energies determined from the cohesive energies of graphene 
and three different graphynes considered here.      
}
\end{abstract}

\pacs{71.15.Nc, 02.70.Ss, 61.48.Gh, 62.23.Kn}

\maketitle


\section{Introduction} 
Graphite and diamond had been only two stable carbon structures found in nature until various types 
of low-dimensional carbon allotropes, including fullerenes, carbon nanotubes, and graphene, 
were recently discovered. For the past decades, these carbon-based nanomaterials have attracted 
a great deal of attention because of their unique electronic and mechanical properties. 
Among them, graphene, a two-dimensional (2D) atomic layer of $sp^2$-bonded carbon atoms, 
exhibits a Dirac cone structure near the Fermi surface, causing its charge carriers 
to behave as massless Dirac fermions, and possesses pseudospin degrees of freedom~\cite{novoselov05,geim07,geim09}. These features of graphene have prompted active investigation 
of its possible applications for novel nanoelectronic and spintronic devices~\cite{williams07,min12}. 
Because of its large surface area and high electrical conductivity, graphene has also been studied 
for a new energy storage material. While graphite, stacks of graphene sheets, is currently used 
as an anode material for Li-ion battery (LIB) because of its high Coulombic efficiency~\cite{schalkwijk02},
there have been intensive studies, both theoretically and experimentally, on the potential of graphene 
as a high-capacity LIB anode material~\cite{yoo08,liang09,wang09,pan09,guo09,wang09-2,lian10,bhardwaj10}. 

Some years ago, graphynes, $sp$-$sp^2$ hybridized networks of carbon atoms, were proposed 
as possible 2D carbon structures with larger surface area than graphene~\cite{baughman87}. 
Recent density-functional theory (DFT) calculations showed that graphynes could have 
some intriguing electronic properties similar to those of graphene, {\it{e.g.}}, 
Dirac cones~\cite{malko12,kim12} and very high carrier mobilities~\cite{chen13}. Furthermore, 
it was predicted that multilayer graphynes could be employed for high-capacity hydrogen storage
materials~\cite{hwang12} and for effective LIB anode materials with higher Li diffusivity~\cite{zhang11} 
and higher Li-storage capacity~\cite{hwang13} than graphite, if they can be produced.
 
The DFT methods have been successfully applied to understand electronic and mechanical characteristics 
of newly-synthesized nanomaterials and even to design novel materials with some intriguing physical 
properties without any experimental evidence. However, it has some well-known limitations; 
for example, the DFT calculations based on the exchange-correlation potential within the Local Density
Approximation (LDA) tends to overestimate the cohesive energy of a solid. 
Even with much experimental effort motivated by the promising DFT-based predictions on graphynes, 
only flakes or building blocks of finite-size graphynes have been synthesized 
so far~\cite{tobe03,haley08,diederich10} but there has been no successful report yet 
for fabrication of extended 2D graphynes. The experimental difficulty in the synthesis of graphynes 
leads us to speculate that the DFT prediction on graphyne energetics may be too optimistic 
as a result of an intrinsic limitation of the methodology. The relative energetic stabilities 
of these new carbon-based nanomaterials need to be analyzed through accurate numerical calculations 
which allow full incorporation of the electron-electron correlation. 
 
Here we performed quantum Monte Carlo (QMC) simulations to investigate ground-state energetics 
of various carbon structures such as diamond, graphene, graphynes, and carbynes, 
especially their cohesive energies. It is found that the cohesive energies of these carbon allotropes 
are significantly overestimated in the DFT calculations when compared with the corresponding QMC energies, 
as many previous studies have reported~\cite{spanu09,hood12}. We also find that the accuracy 
of the DFT cohesive energy depends on its carbon-carbon bonding nature; the discrepancy 
between the QMC cohesive energy and the DFT energy is the smallest for $sp^3$-bonded diamond, 
and the largest for carbyne, one-dimensional (1D) chain of $sp$-bonded carbon atoms. 
In the case of graphene and graphynes, the magnitude of the overestimation of the DFT cohesive energy 
increases systematically as the ratio of the $sp$-bonded carbon atoms increases. According to our QMC calculations, the cohesive energy of graphene is 7.464(10) eV/atom, which is larger 
than the cohesive energy of $\gamma$-graphyne, energetically the most stable graphyne structure, 
by 0.698(12) eV/atom.
 
\section{Methods and Computational Details}
\label{sec:methods}
The QUANTUM ESPRESSO package~\cite{giannozzi09} was used for the DFT calculations where we employed 
the exchange correlation potential with the generalized gradient approximation in the 
Perdew-Burke-Ernzerhof parameterization (PBE) scheme~\cite{perdew96}. 
The energy-consistent pseudopotential (PP) proposed by Burkatzki, Filippi, 
and Dolg (BFD)~\cite{burkatzki07,burkatzki08}, was used for the valence-core partitioning 
in describing a carbon atom with the kinetic energy cutoff of 120 Ry. The first Brillouin zone integration 
was done using the Monkhorst-Pack scheme~\cite{monkhorst76}. To avoid spurious interactions 
between image structures due to the periodic calculations, a vacuum layer of 10~\AA~was put 
in each of all non-periodic directions.
  
The QMC calculations were performed using the QMCPACK package~\cite{kim12-2}, 
where we employed trial many-body wave functions of the Slater-Jastrow (SJ) type 
with a single Slater determinant consisting of the Kohn-Sham orbitals obtained from the DFT calculations. 
A variant of the linear method of Umrigar and coworkers~\cite{umrigar07} was used in the variational 
Monte Carlo calculations to optimize the variational parameters in one-body Jastrow factors as well as 
in two-body ones. Then the diffusion Monte Carlo (DMC) method was applied to compute 
the exact ground-state energies within the fixed-node approximation~\cite{foulkes01}. 
Twist-averaged boundary conditions (TABC) were employed to minimize one-body finite-size effects 
in the supercell calculations~\cite{lin01}. For diamond, the modified potential~\cite{drummond08} 
and the Chiesa~\cite{chiesa06} corrections are additionally applied for the two-body finite-size effects.
Finally the TABC energies of the supercells of various sizes were extrapolated to the thermodynamic limit 
to obtain the ground-state energy of the corresponding bulk solid. These DMC procedures were 
rigorously applied in Ref.~\cite{shulenburger13} to study the bulk properties of various materials 
with different bonds.
 
\section{Results}
\label{sec:results}
Using the experimental values for the carbon-carbon bond length (1.421~\AA~for 
graphene or graphite~\cite{lynch66} and 1.545~\AA~for diamond~\cite{hom75}), 
we first performed the DFT-PBE calculations on graphene and diamond to compute their cohesive energies, 
which is defined by $E_{\text{coh}} = (N_{\text{atom}} E_{\text{atom}}-E_{\text{tot}})/N_{\text{atom}}$ 
with $N_{\text{atom}}$, $E_{\text{atom}}$ and $E_{\text{tot}}$ being the number of carbon atoms 
per unit cell, the total energy of a carbon atom in a vacuum, and the total energy of graphene 
(or diamond) per unit cell, respectively. With the 12 x 12 x 1 (12 x 12 x 12) k-point sampling, 
the PBE-PP cohesive energy was found to be 7.906 eV/atom (7.778 eV/atom) for graphene (diamond). 
This value for graphene is consistent with the cohesive energy of 7.972 eV/atom computed 
with the projector-augmented-wave (PAW) method as implemented in Vienna Ab-initio Simulation Package (VASP) 
and is also comparable to the previously-reported DFT cohesive energies of 7.9 eV/atom~\cite{koskinen08} 
and 7.73 eV/atom~\cite{ivanovskaya10}. This tells us that the valence-core partitioning of 
the energy-consistent BFD pseudo-potential used here is accurate in describing the cohesion of carbon atoms.  
 
\begin{figure}[t]
\vspace{-0.3cm}
\includegraphics[width=4in]{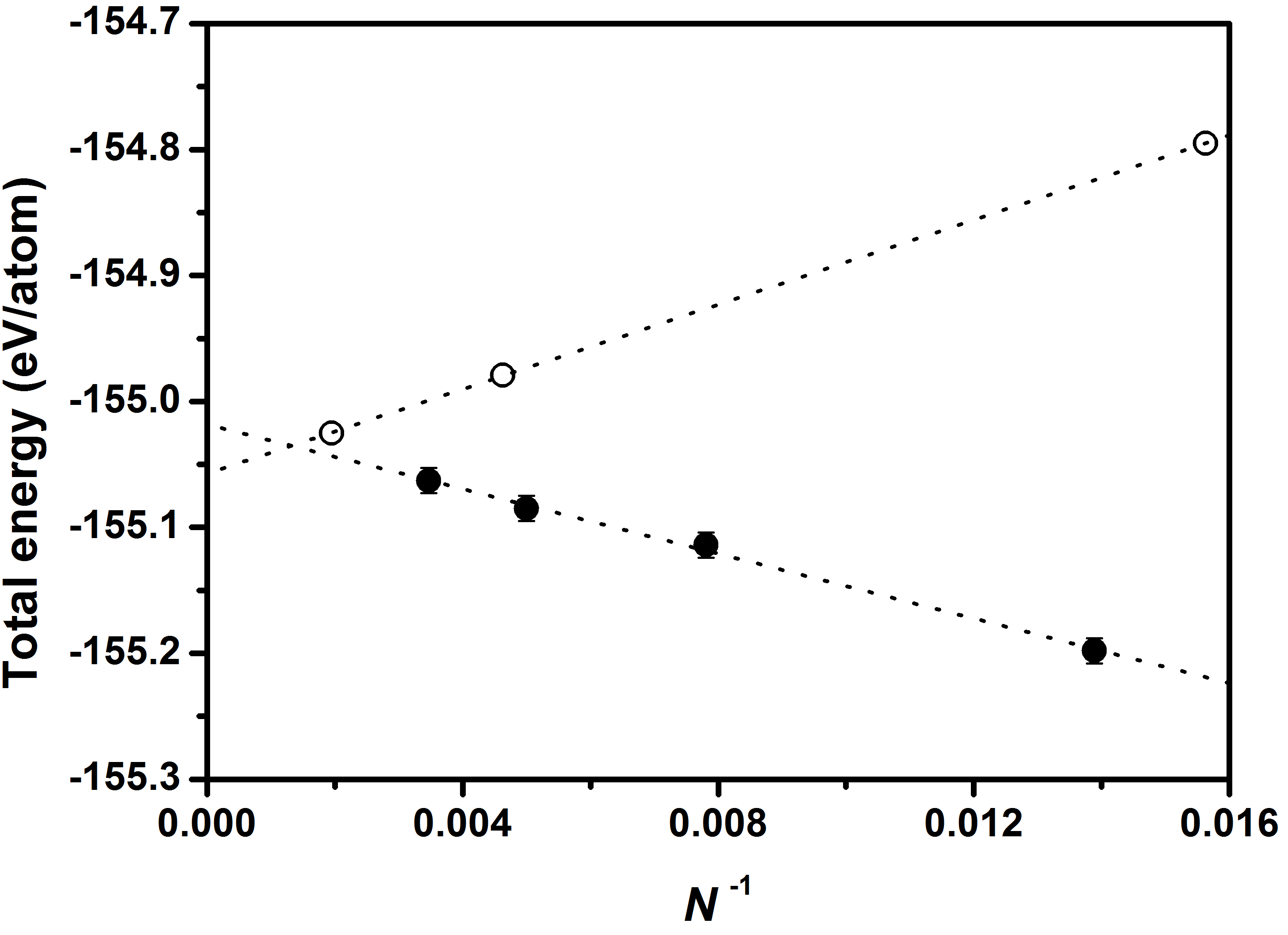}
\vspace{-0.2cm}
\caption[0]{Twist-averaged DMC energies per atom of graphene (solid circles) and diamond (open circles) supercells of various sizes as a function of 1/$N$ with $N$ being the number of electrons per supercell. 
The dotted lines represent simple-linear-regression fits. The statistical errors are smaller 
than the sizes of the symbols. 
}
\label{fig:diamond}
\end{figure}

We then carried out the QMC calculations on graphene and diamond. Various sizes of supercells, 
up to the 6 x 6 x 1  (72 carbon atoms) supercell for graphene and the 4 x 4 x 4 (128 carbon atoms) supercell
for diamond, were simulated with twisted boundary conditions which were designed to include 
all k-points sampled in the DFT calculations for the respective material. Figure~\ref{fig:diamond} shows 
the TABC-DMC energies per atom for graphene and diamond supercells with respect to the inverse 
of the number of electrons per supercell. From the simple-linear-regression fits the ground-state energies 
in the thermodynamic limit were estimated to be -155.018(8) eV/atom for graphene and -155.057(1) eV/atom 
for diamond. With the same BFD pseudo-potential used here, the DMC energy of a single carbon atom was 
computed to be -147.554(3) eV. From this, the DMC cohesive energies of graphene and diamond were estimated 
to be 7.464(10) eV/atom and 7.503(3) eV/atom, respectively. We tried a linear fit 
of the graphene supercell energies as a function of $N^{-5/4}$, which corresponds to the size dependence 
of supercell energies of a strictly two-dimensional electron system~\cite{drummond08}. 
This also produced an excellent fit with the extrapolated value of -155.037(4) eV/atom in the bulk limit, 
lower only by 0.019(9) eV/atom than the corresponding value from the $N^{-1}$ linear fit. 
For further discussion, we made the finite-size analysis of the DMC supercell energies with the fit 
to a linear function of $N^{-1}$ for all carbon structures, regardless of their dimensionalities, 
and ignored higher-order corrections to the volume~\cite{drummond08}.
 
The PBE and DMC cohesive energies of graphene and diamond, along with their zero-point energy (ZPE) corrections, are summarized in Table~\ref{tbl:gradiaE}, where the DMC cohesive energies of graphite were estimated through the addition of the previously-reported DMC interlayer binding energy~\cite{spanu09} 
to the corresponding energies of a single graphene sheet. As seen in Table~\ref{tbl:gradiaE}, 
the PBE cohesive energies are significantly larger than the DMC energies in both graphene and diamond,
indicating that the DFT calculations based on the PBE exchange-correlation potential tend 
to overestimate the carbon binding as in LDA calculations. The DMC cohesive energy of graphite is found 
to be slightly larger than that of diamond, which is consistent with the fact that graphite is 
energetically the most stable carbon structure. The ZPE-corrected DMC cohesive energies of both graphite 
and diamond are in excellent agreement with the corresponding experimental values~\cite{brewer}. 
For the cohesive energy difference between graphite and diamond, the DMC result of $\sim$27 meV/atom 
is basically identical to the experimentally-reported value. This, largely due to the error cancellations, 
is fortuitous given that there are remaining systematic errors in our QMC calculations, 
{\it{e.g.}}, fixed-node errors and higher-order finite-size corrections. More comprehensive calculations 
beyond the SJ orbitals for larger supercells are called for to resolve the energy differences 
within 10 meV. Nevertheless, the results support the use of the QMC methods to compute 
accurate relative energetics of different carbon allotropes. 
 
\begin{figure}[t]
\vspace{-0.3cm}
\includegraphics[width=6in]{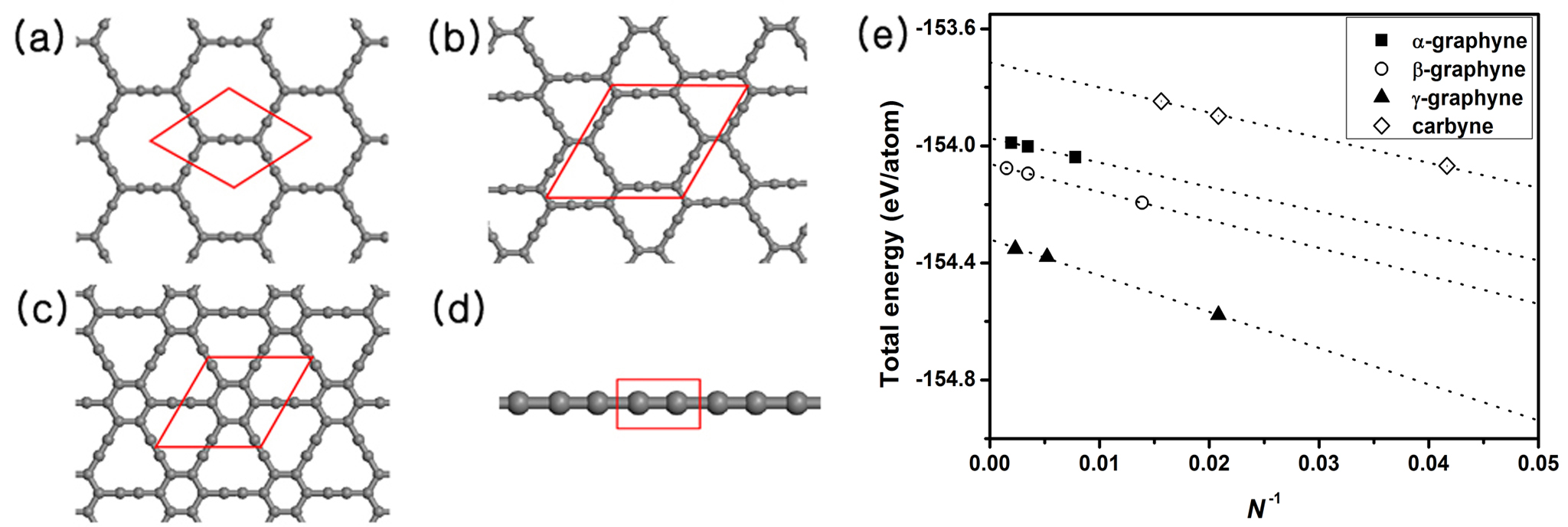}
\vspace{-0.2cm}
\caption[0]{(Color online) Atomic structures of (a) $\alpha$-, (b) $\beta$-, (c) $\gamma$-graphyne, and (d) carbyne, whose unit cells are denoted by the red parallelograms. The grey dots represent the carbon atoms. (e) Twist-averaged DMC total energy per atom of the graphyne and the carbyne supercells of various sizes as a function of the inverse of the total number of electrons per supercell. The dotted lines represent simple-linear-regression fits. The statistical errors are smaller than the sizes of the symbols.
}
\label{fig:graphy}
\end{figure}

The DFT and the QMC calculations were performed for recently-proposed low-dimensional carbon allotropes 
of graphynes and carbyne, whose atomic structures are shown in Fig.~\ref{fig:graphy}(a)-(d), 
to analyze their energetic stabilities relative to that of graphene. The geometry optimization 
was carried out in the DFT calculation for each of these structures until the Hellman-Feynman force 
acting on a carbon atom was less than 10$^{-2}$ eV/\AA, which resulted in the carbon-carbon bond lengths 
of three graphynes identical to the previously-reported values~\cite{kim12} with errors less than 0.2\%. 
For carbyne, we here consider a linear carbon chain of alternating single and triple bonds (polyyne), 
which is known to be more stable than a chain of only double bonds (polycumulene)~\cite{eisler05,luo09}. 
While the PBE total energies of graphynes (carbyne) were computed with the 12 x 12 x 1 (12 x 1 x 1) 
k-point sampling, the twist boundary conditions were applied in the QMC calculations to minimize 
the size effects. Figure~\ref{fig:graphy}(e) shows the TABC-DMC energies per atom of the graphyne 
and carbyne supercells as a function of the inverse of the number of electrons per supercell. 
As in the QMC calculations for graphene and diamond, the TABC-DMC energies are in excellent linear fits 
with the inverse of the supercell sizes, which are represented by the dotted lines of 
Fig.~\ref{fig:graphy}(e). Substracting the TABC-DMC energies extrapolated to the bulk limit 
from the DMC atomic energy of carbon, we estimated the DMC cohesive energies of these nanomaterials, 
which are shown in Table~\ref{tbl:graphyE} along with their PBE cohesive energies based 
on two different carbon pseudopotentials. Here PBE-PP represents the PBE cohesive energies computed 
with the energy-consistent BFD pseudopotential while PBE-PAW corresponds to the energies calculated 
using the PAW method implemented in the VASP. 
 
As shown in Table~\ref{tbl:graphyE}, the PBE-PP cohesive energies are found to be smaller 
than the corresponding PBE-PAW energies, by 46 to 75 meV/atom, depending on the specific carbon structure.
These discrepancies are much smaller than the differences between the DMC cohesive energies 
and either of the PBE energies, which are shown to be as large as $\sim 800$ meV/atom for carbyne. 
Noting that the DMC calculations were done with the BFD pseudopotential, we analyze the accuracy 
of the PBE calculations for these low-dimensional carbon structures by comparing the PBE-PP cohesive 
energies with the corresponding DMC energies. The PBE calculations significantly overestimate 
the cohesive energies of 2D graphynes and 1D carbine and the degree of the PBE overestimation is seen 
to vary as the carbon-carbon bonding nature changes.
 
\begin{figure}[t]
\vspace{-0.3cm}
\includegraphics[width=4in]{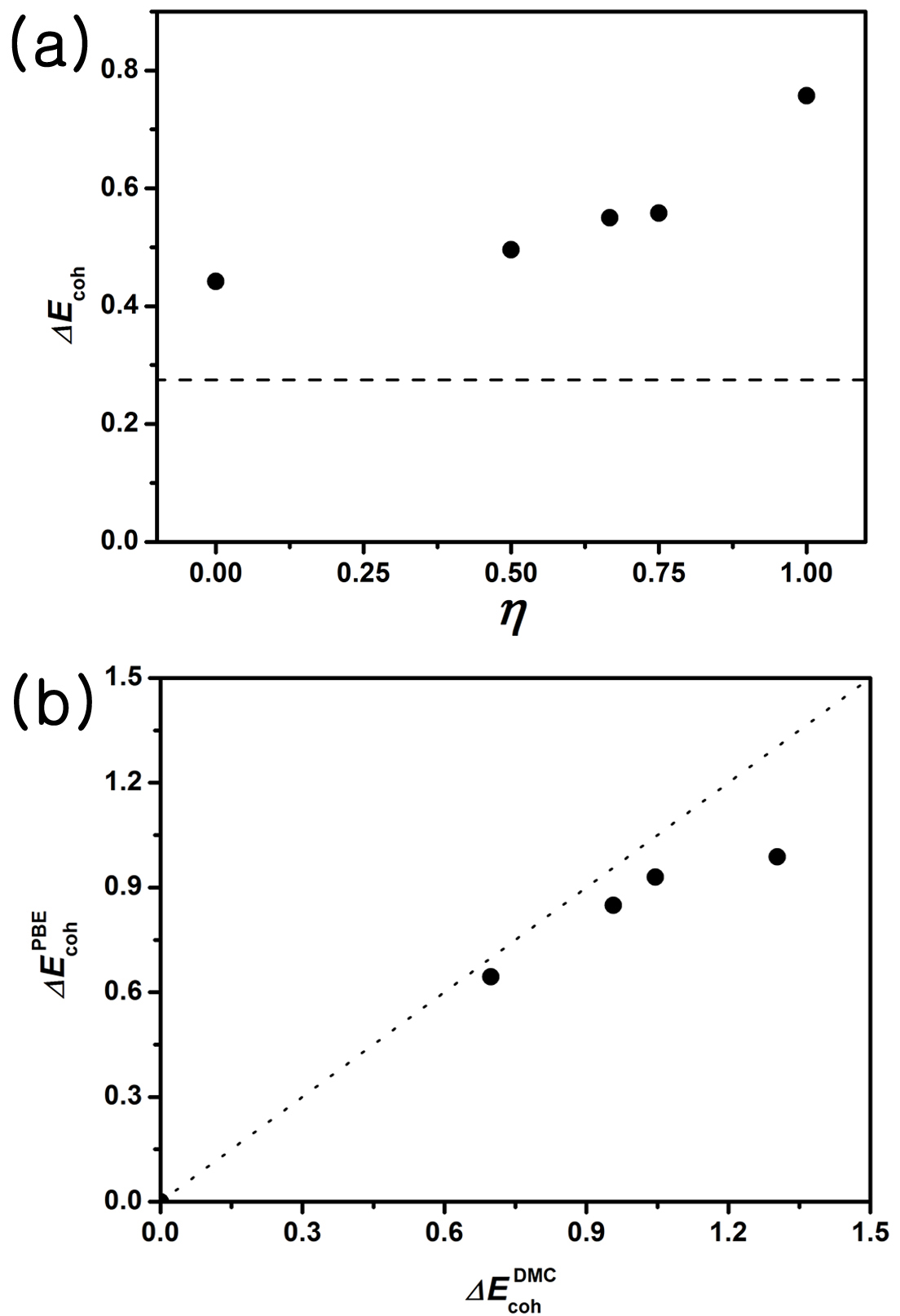}
\vspace{-0.2cm}
\caption[0]{(a) Difference between the DMC cohesive energy and the PBE-PP energy as a function of $\eta$, the ratio of $sp$-bonded carbon atoms, for several carbon allotropes and (b) the PBE-PP cohesive energy of a carbon allotrope relative to graphene($\Delta E_{\text{coh}}^{\text{PBE}}$) versus its DMC cohesive energy relative to graphene ($\Delta E_{\text{coh}}^{\text{DMC}}$). The dashed horizontal line in (a) represents the PBE-DMC cohesive energy difference for $sp^3$-bonded diamond and the dotted line in (b) corresponds to $\Delta E_{\text{coh}}^{\text{PBE}}$=$\Delta E_{\text{coh}}^{\text{DMC}}$. The energies are in units of eV/atom.
}
\label{fig:deltacoh}
\end{figure}

Figure~\ref{fig:deltacoh}(a) shows the differences between the DMC cohesive energies of 
several carbon allotropes and their PBE-PP cohesive energies as a function of the ratio of $sp$-bonded 
carbon atoms, $\eta=N_{sp}/N_{\text{atom}}$ where $N_{sp}$ and $N_{\text{atom}}$ are the numbers 
of $sp$-bonded carbon atoms and of all carbon atoms in a unit cell, respectively. One can see that 
the PBE-DMC cohesive energy differences for $sp$-$sp^2$ hybridized 2D graphynes and $sp$-bonded 1D carbyne 
are larger than the corresponding value of 0.458(10) eV/atom for graphene. As a matter of fact, 
the cohesive energy difference increases monotonically as $\eta$ increases from zero (graphene) 
to one (carbyne). Note that $\eta$ is 0.5, 0.67, and 0.75 for $\gamma$, $\beta$, and $\alpha$-graphyne,
respectively. This leads us to conclude that more correlation energies were missed in the PBE calculations 
for electrons participating in the $sp$ bonds than for those in the $sp^2$ bonds. The PBE-DMC cohesive 
energy difference shows the largest value in 1D $sp$-bonded carbyne while the 3D $sp^3$-bonded diamond 
has the smallest energy difference (see the dashed horizontal line of Fig.~\ref{fig:deltacoh}(a)), 
indicating that the dimensionalities of carbon structures as well as their bonding nature affect 
the accuracy of the PBE calculations for their cohesive energies. Figure~\ref{fig:deltacoh}(b) shows 
the PBE cohesive energy of a carbon allotrope relative to that of graphene, 
$\Delta E_{\text{coh}}^{\text{PBE}}=E_{\text{coh}}^{\text{PBE}}(\text{graphene}) 
- E_{\text{coh}}^{\text{PBE}}(\text{allotrope})$, versus the relative DMC cohesive energy, 
$\Delta E_{\text{coh}}^{\text{DMC}} = E_{\text{coh}}^{\text{DMC}}(\text{graphene}) - E_{\text{coh}}^{\text{DMC}}(\text{allotrope})$. 
As can be seen, the relative PBE cohesive energies are smaller 
than the corresponding DMC energies in all low-dimensional allotropes considered here. 
From this we conclude that these new carbon-based nanomaterials are energetically not as stable 
as the PBE calculations have predicted.
 
\begin{figure}[t]
\vspace{-0.3cm}
\includegraphics[width=4in]{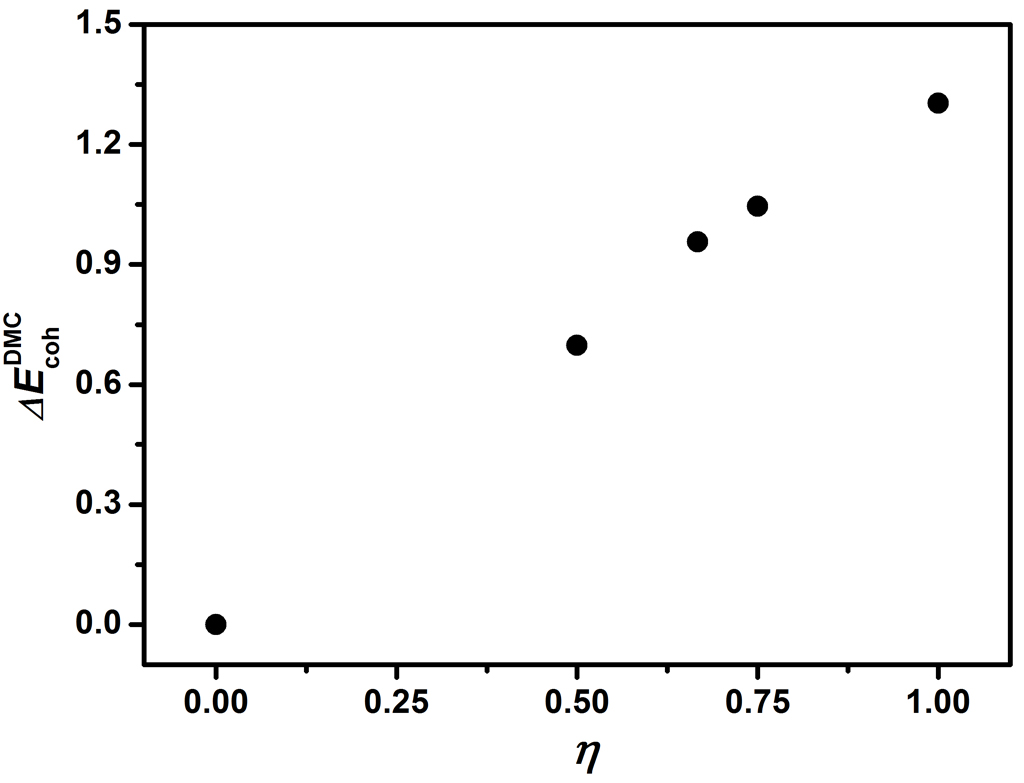}
\vspace{-0.2cm}
\caption[0]{The DMC cohesive energy difference between a low-dimensional carbon allotrope and graphene, $\Delta E_{\text{coh}}^{\text{DMC}}$=$E_{\text{coh}}^{\text{DMC}}$(graphene)-$E_{\text{coh}}^{\text{DMC}}$(allotrope) , as a function of the ratio of $sp$-bonded carbon atoms $\eta$. The energies are in units of eV/atom.
}
\label{fig:DMCdeltacoh}
\end{figure}

The DMC cohesive energy difference between a low-dimensional carbon allotrope and graphene is shown 
in Fig.~\ref{fig:DMCdeltacoh} as a function of $\eta$, the ratio of the $sp$-bonded carbon atoms. 
According to our DMC calculations, $\gamma$-graphyne ($\eta$=0.5), which is the most stable 
among new carbon structures presented in Fig.~\ref{fig:DMCdeltacoh}, has smaller cohesive energy 
than graphene by 0.698(12) eV/atom, while the cohesive energy of carbyne is smaller by 1.303(11) eV/atom 
than that of graphene. Experimental difficulty encountered in the synthesis of graphynes and carbynes 
could be explained by their significantly smaller cohesive energies than graphene. It is seen that 
the cohesive energies of $sp$-$sp^2$ hybridized graphynes decrease monotonically as $\eta$ increases. 
This is consistent with the fact that the $sp$-bonded carbon atoms are bound only through two bonds, 
either one single bond and one triple bond or two double bonds, while the $sp^2$-bonded atoms are 
bound through three bonds. Figure~\ref{fig:DMCdeltacoh} gives a guideline for the energetic stability 
of a $sp$-$sp^2$ hybridized 2D carbon network relative to graphene.
 
As noted some years ago by Baugman {\it{et al.}}~\cite{baughman87}, one can design many different types 
of graphynes that are 2D carbon phases involving acetylene bonds. Among them, (6,6,12)-graphyne 
with a rectangular symmetry has recently drawn much interest because it possesses direction-dependent 
Dirac cones to allow electron collimation transport without any external field~\cite{malko12}. 
Quantitative prediction for the cohesion energetics of a new graphyne structure can be made 
by analyzing the contribution of each carbon bond to their cohesive energies. 
There are three different carbon-carbon bonds involved in a 2D hybridized graphyne structure, 
that is, a single bond (C-C), a double bond (C=C) and a triple bond (C$\equiv$C). 
Its cohesive energy can be assumed to be the sum of the bond energies;
\begin{equation}
N_{\text{atom}}E_{\text{coh}}=\varepsilon_{\text{s}}N_{\text{s}}
+\varepsilon_{\text{d}}N_{\text{d}}+\varepsilon_{\text{t}}N_{\text{t}},
\label{eq:bondE}
\end{equation}
where $E_{\text{coh}}$ is the cohesive energy per atom and $\varepsilon_{\text{s}}$, $\varepsilon_{\text{d}}$
and $\varepsilon_{\text{t}}$ represent the single, the double, and the triple bond energies, respectively. 
Here $N_{\text{atom}}$, $N_{\text{s}}$, $N_{\text{d}}$, and $N_{\text{t}}$ are the numbers 
of the carbon atoms, the single bonds, the double bonds, and the triple bonds per unit cell, respectively. 
By fitting the cohesive energies of graphene and three different graphyne structures 
in Table~\ref{tbl:graphyE} to Eq.~(\ref{eq:bondE}), we obtained the C-C, the C=C, and the C$\equiv$C 
bond energies of 2D hybridized carbon networks. As shown in Table~\ref{tbl:bondE}, 
the multiple-linear-regression fits of the DMC and both PBC data resulted in the correlations coefficients 
very close to the unit value, justifying our assumption that the cohesive energy of a graphyne structure 
can be described by the sum of the carbon-carbon bond energies. With the PBE-PAW bond energies 
in Table~\ref{tbl:bondE}, the cohesive energy of (6,6,12)-graphyne is estimated to be 7.265(22) eV/atom, 
which is in excellent agreement with its PBE-PAW cohesive energy of 7.264 eV/atom. 
This strongly validates our approach to the cohesive energy with the carbon-carbon bond energies. 
We expect that with the DMC bond energies in Table~\ref{tbl:bondE}, one can make an accurate prediction 
for the cohesive energy of a newly-proposed graphyne. The predicted DMC cohesive energy of (6,6,12)-graphyne 
is 6.671(31) eV/atom. 

\section{Conclusions}
\label{sec:conclusions}
The cohesion energetics of various carbon allotropes have been studied using the QMC methods 
with full description of electron-electron correlation. We have found that the DFT calculations 
based on the PBE exchange-correlation potential significantly overestimate the cohesive energies 
of these carbon-based materials. The degree of the PBE overestimation is found to increase as the ratio 
of $sp$-bonded carbon atoms increases. This suggests that the PBE predictions for the energetic stabilities 
of $sp$-$sp^2$ hybridized 2D carbon networks such as graphynes, relative to that 
of $sp^2$-bonded graphene, should be considered with a great deal of caution. It is also found that 
the DMC cohesive energies of graphite and diamond are in agreement with the corresponding experimental 
values reported some years ago. The DMC results for the cohesive energies of several types of graphynes 
have revealed their systematic decrease with the increase of the ratio of $sp$-bonded atoms. 
Finally we conclude that accurate quantitative prediction for the cohesion energetics 
of a newly-proposed graphyne can be made using the carbon-carbon bond energies determined 
from the DMC cohesive energies of graphene and three different 
($\alpha$-, $\beta$-, and $\gamma$-) graphynes.
 
\begin{acknowledgments}
This work was supported by the Basic Science Research Program (2012R1A1A 2012006887) 
through the National Research Foundation of Korea funded by the Ministry of Education, Science 
and Technology, and by Materials Sciences and Engineering Division, Office of Basic Energy Sciences, 
U.S. Department of Energy. We also acknowledge the support from the Supercomputing Center/Korea Institute 
of Science and Technology Information with supercomputing resources including technical support 
(KSC-2013-C3-033).
\end{acknowledgments}

\bibliography{cohesive}

\newpage

\begin{table}[t] 
\caption{Cohesive energies of graphene, graphite, and diamond in units of eV/atom. 
The zero-point energy corrections used here are 0.166 eV/atom for graphene and graphite~\cite{aljishi82,hasegawa04} 
and of 0.176 eV/atom for diamond~\cite{schimka11} 
and the experimental values are from Ref.~\cite{brewer}. 
The DMC cohesive energy for graphite is obtained by adding the previously-reported DMC interlayer 
binding energy~\cite{spanu09} of 0.056(5) eV/atom to the cohesive energies of a single graphene sheet.}
\begin{tabular}{ >{\centering}m{1in} | >{\centering}m{1in} | >{\centering}m{1in} | >{\centering}m{1in} | >{\centering}m{1in} | >{\centering}m{1in} }   
       \hline \hline
      & PBE & PBE+ZPE & DMC & DMC+ZPE  & Exp. \tabularnewline \hline
     Graphene & 7.906 & 7.740 & 7.464(10) & 7.298(10)  & - \tabularnewline \hline
     Graphite & - & - & 7.520(11) & 7.354(11) & 7.374 \tabularnewline \hline
     Diamond  & 7.778 & 7.602 & 7.503(3) & 7.327(3)  & 7.346  \tabularnewline
     \hline \hline
\end{tabular}
\label{tbl:gradiaE}
\end{table}
 
\clearpage
  
\begin{table}[t] 
\caption{Cohesive energies of low-dimensional carbon allotropes in units of eV/atom.} 
   \begin{tabular}{ >{\centering}m{1in} | >{\centering}m{1in} | >{\centering}m{1in} | >{\centering}m{1in} | >{\centering}m{1in} | >{\centering}m{1in} }
     \hline \hline
      & graphene & $\gamma$-graphyne & $\beta$-graphyne & $\alpha$-graphyne  & carbyne \tabularnewline \hline
     DMC & 7.464(10) & 6.766(6) & 6.507(3) & 6.418(3)  & 6.161(5) \tabularnewline \hline
     PBE-PP & 7.906 & 7.262 & 7.057 & 6.976 & 6.918\tabularnewline \hline
     PBE-PAW & 7.972 & 7.330 & 7.129 & 7.051  & 6.964  \tabularnewline
     \hline \hline
   \end{tabular}  
\label{tbl:graphyE}
\end{table}
 
\clearpage

\begin{table}[t] 
\caption{Carbon-carbon bond energies$^a$ of $sp$-$sp^2$ hybridized graphyne structure in units of eV 
which were determined through multiple-linear-regression fits of the cohesive energies of graphene 
and three different graphyne structures shown in Table~\protect\ref{tbl:graphyE} to Eq. (1). }  
   \begin{tabular}{ >{\centering}m{1in} | >{\centering}m{1in} | >{\centering}m{1in} | >{\centering}m{1in} | >{\centering}m{1.5in} }
     \hline \hline
      & C-C & C=C & C$\equiv$C & correlation coefficient \tabularnewline \hline
     DMC & 4.403(33) & 6.143(42) & 7.652(59) & 0.99999 \tabularnewline \hline
     PBE-PP & 4.520(17) & 6.735(24) & 8.762(32) & 1.00000 \tabularnewline \hline
     PBE-PAW & 4.537(23) & 6.820(33) & 8.900(43) & 1.00000  \tabularnewline
     \hline \hline
   \end{tabular}\\
\label{tbl:bondE}
\end{table}
  
\clearpage

\clearpage

\clearpage

\clearpage

\end{document}